\documentclass[aps,prd,twocolumn,showpacs,showkeys,floatfix,nofootinbib,superscriptaddress]{revtex4-1}
\usepackage{dcolumn}

\usepackage{amsmath,amsfonts,amssymb,mathrsfs,times,bm}
\usepackage{extarrows}
\usepackage{changes}
\usepackage{graphicx}
\usepackage{float}
\usepackage{graphicx,epstopdf}
\usepackage{dcolumn}
\usepackage{exscale}
\usepackage{relsize}
\usepackage{mathtools}
\usepackage{soul}
\usepackage[colorlinks=true,breaklinks=true,linkcolor=blue,citecolor=blue,urlcolor=blue]{hyperref}
\newcommand{\nn}{\nonumber}

\begin{document}

\title[]
{Finite-distance gravitational deflection of massive particles by the Kerr-like black hole in the bumblebee gravity model}
\author{Zonghai Li}
\email{lzh@my.swjtu.edu.cn}
\affiliation{Center for Theoretical Physics, School of Physics and Technology, Wuhan University, Wuhan, 430072, China}

\author{Ali {\"O}vg{\"u}n}
\email{ali.ovgun@pucv.cl}
\affiliation{Instituto de F{\'i}sica, Pontificia Universidad Cat{\'o}lica de
Valpara{\'i}so, Casilla 4950, Valpara{\'i}so, Chile.}
\affiliation{Physics Department, Arts and Sciences Faculty, Eastern Mediterranean
University, Famagusta, North Cyprus via Mersin 10, Turkey.}
\date{\today}

\begin{abstract}
In this paper, we study the weak gravitational deflection angle of relativistic massive particles by the Kerr-like black hole in the bumblebee gravity model. In particular, we focus on weak field limits and calculate the deflection angle for a receiver and source at a finite distance from the lens. To this end, we use the Gauss-Bonnet theorem of a two-dimensional surface defined by a generalized Jacobi metric. The spacetime is asymptotically non-flat due to the existence of
a bumblebee vector field. Thus the deflection angle is modified and can be divided into three parts: the surface integral of the Gaussian curvature, the path integral of a geodesic curvature of the particle ray and the change in the coordinate angle. In addition, we also obtain the same results by defining the deflection angle. The effects of the Lorentz breaking constant on the gravitational lensing are analyzed. In particular, we correct a mistake in the previous literature. Furthermore, we consider the finite-distance correction for the deflection angle of massive particles.



\end{abstract}

\maketitle

\section{Introduction}
In physics, fundamental interactions can be described mathematically by field theories. Gravitation is defined as a classic field in curved spacetime by Einstein's general relativity (GR). The other three interactions are described as quantum fields by the standard model (SM) of particle physics. GR and SM have an intersection at the Planck scale ($\sim10^{19}$ GeV), and modern physics is committed to unifying these two theories. For this purpose, some quantum gravitational theories have been proposed; however, it is currently impossible directly to test these theories through experimentation. Fortunately, some signals of quantum gravity can emerge at sufficiently low energy scales, and their effects can be observed in experiments carried out at current energy levels. One of these signals may be associated with the breaking of Lorentz symmetry~\cite{Casana2018}.

In 1989, Kosteleck\'{y} and Samuel~\cite{KS1989} proposed the bumblebee gravity theory as the simplest model for studying the spontaneous Lorentz symmetry breaking (LSB), in which a bumblebee field with a vacuum expectation value leads to spontaneous breaks in Lorentz symmetry. The interest towards in bumblebee gravity theory has increased over the years~\cite{VAK2004,Bertolami2005,Bailey2006,Bluhm2008,VAK2009,Seifert2010,Maluf2014,Guiomar2014,Santos2015,Maluf2015,Escobar2017,JFA2019,Jesus2019}. In particular, some new exact solutions have been recently found. In 2018, Casana \textit{et al} obtained an exact Schwarzschild-like black hole solution in this bumblebee gravity model and then studied it with some classical tests~\cite{Casana2018}. The gravitational deflection angle of light~\cite{Ali181} and the Hawking radiation~\cite{Kanzi2019} in this Schwarzschild-like spacetime were also studied recently. Earlier in 2019, \"{O}vg\"{u}n \textit{et al} adroitly obtained a traversable wormhole solution that can deduce the Ellis wormhole~\cite{Ali2019} and studied the gravitational lensing of light. Within the same time-frame, Ding \textit{et al} found a Kerr-like solution in this bumblebee gravity model and studied its shadow and accretion disk~\cite{Ding2019,Liu2019}.

On the other hand, as a powerful tool of astrophysics and cosmology, the gravitational lensing has been used to test the fundamental theory of gravity~\cite{DED1920,Will2015}, to measure the mass of galaxies and clusters ~\cite{Hoekstra2013,Brouwer2018,Bellagamba2019}, to detect dark matter and dark energy~\cite{Vanderveld2012,cao2012,zhanghe2017,Huterer2018,SC2019,Andrade2019}, and so on. The analysis of the signatures of the gravitational lensing of light or massive particles may be useful in testing the bumblebee gravity model and detecting the LSB effects. Thus, this paper will study the gravitational lensing in bumblebee gravity model.

Recently, Gibbons and Werner proposed a geometrical and topological method of studying weak gravitational lensing~\cite{GW2008,Werner2012}. In this method, the Gauss-Bonnet (GB) theorem is applied to the corresponding optical geometry and the deflection angle is calculated by integrating the Gaussian curvature of the optical metric. In particular, the Gibbons-Werner method shows that the deflection angle can be viewed as a global topological effect. By using Gibbons-Werner method, many studies have explored the gravitational deflection angle of light from different lens objects within the different gravitational theories~\cite{Jus172,  Ovgun:2018prw,Ovgun:2018fte,Jusufi:2018kmk,Ovgun:2019qzc,Jus181,Ovgun:2018tua,Jus183,Ali182,Jusufi:2017vew,Goulart2018,Sakalli2017,Jus2016,Kumaran:2019qqp,Ovgun:2018oxk,Javed2019,Leon2019,LZ2019,Jus174,Jus184,Jus187}. Furthermore, some authors utilized this method to study the deflection angle of light from asymptotically  non-flat gravitational sources~\cite{Jus175,Jus185,Ali181,Ali191,Jusufi:2017vta}, the deflection angle of light in a plasma medium~\cite{CG2018,CGV2019}, and the deflection angles of neutral and charged massive particles~\cite{CG2018,CGV2019,LHZ2019,jaka}.

By using the GB theorem, some authors were able to compute the finite-distance gravitational deflection angle of light. In this case, the receiver and source are assumed to be at finite distance from a lens object, which is different from the usual consideration where receiver and source are at infinite distance from the lens. First, Ishihara \textit{et al} used the GB theorem to study the finite-distance deflection of light in static and spherically symmetric spacetime~\cite{ISOA2016,IOA2017}. Then, Ono \textit{et al} proposed the generalized optical metric method and investigated the finite-distance deflection angle of light in stationary, axisymmetric spacetime~\cite{OIA2017,OIA2018,OIA2019}. In these studies, the possible astronomical application according to finite-distance correlations was considered. As well, Ono and Asada gave a comprehensive review on finite-distance deflection of light \cite{OA2019}. In addition, Arakida applied different definitions of the deflection angle to study the finite-distance deflection of light \cite{Arakida2018}. Very recently, Crisnejo and Gallo considered the finite-distance deflection of light in a spherically symmetric gravitational field with a plasma medium~\cite{CG2019}.

The aim of this paper is to investigate the finite-distance deflection of massive particles by the Kerr-like black hole in the bumblebee gravity model within the weak-field limits. To this end, we will apply the definition of deflection angle given by Ono \textit{et al}~\cite{OIA2017}. In order to use the GB theorem, we shall apply the Jacobi metric method~\cite{Gibbons2016,Chanda2019}. For stationary spacetime, the corresponding Jacobi metric is the Jacobi-Maupertuis Randers-Finsler metric (JMRF).

This paper is organized as follows: In Section~\ref{BG}, we review the Kerr-like black hole solution in the bumblebee gravity theory and then solve the motion equation of massive particles moving on the equatorial plane. In section~\ref{GBDEF}, we study the deflection angle of the massive particles by the Kerr-like black hole for a receiver and source at finite distance using the GB theorem. In Section~\ref{DEFDEF}, we computes the finite-distance deflection angle by definition. Section~\ref{FDC} analyzes the results and considers the finite-distance correction of the gravitational deflection angle of massive particles. Finally, we comment on our results in Section~\ref{CONCLU}. Throughout this paper, we take the unit of $G =\hbar= c = 1$ and the spacetime signature $(-,+,+,+)$.

\section{The kerr-like black hole in the bumblebee gravity model}\label{BG}
\subsection{The black hole solution}
 The bumblebee gravity theory is a typical model of studying the spontaneous Lorentz symmetry breaking. The bumblebee field $B_\mu$ acquires a non-zero vacuum expectation value
 \begin{eqnarray}
<B_\mu>&=&b_\mu~,
\end{eqnarray}
where $b_\mu$ is a constant vector. Its action reads~\cite{Ding2019,Liu2019}
\begin{eqnarray}
\label{action}
S_B&=&\int d^4 x \sqrt{-g} \left[\frac{1}{2\kappa}\bigg(R+\varrho B^\mu B^\nu R_{\mu\nu}\right)\nn\\
&&-\frac{1}{4}B^{\mu\nu}B_{\mu\nu}-V\left( B^{\mu}\right)\bigg]~,
\end{eqnarray}
where $\kappa=8\pi$, $\varrho$ is the coupling constant (with mass dimension $-1$). The bumblebee field strength $B_{\mu\nu}$ and the potential $V$ are defined by
\begin{eqnarray}
&&B_{\mu\nu}=\partial_\mu B_\upsilon-\partial_\nu B_\mu~,\nn\\
&&V\equiv V(B^\mu B_\mu \pm d^2)~.
\end{eqnarray}
where $d^2$ is a positive real constant.
The gravitational field equation in vacuum reads
\begin{eqnarray}
\label{fieldequation}
R_{\mu\nu}&=&\kappa T_{\mu\nu}^{B}+2\kappa g_{\mu\nu}V+\frac{1}{2}\kappa g_{\mu\nu} B^{\alpha\beta} B_{\alpha\beta}-\kappa g_{\mu\nu} B^\alpha B_\alpha V'\nn\\
&&+\frac{\varrho}{4}g_{\mu\nu}\nabla^2(B^\alpha B_\alpha)+\frac{\varrho}{2}g_{\mu\nu}\nabla_\alpha \nabla_\beta(B^\alpha B^\beta)~,
\end{eqnarray}
where $T_{\mu\nu}^{B}$ denotes the bumblebee energy momentum tensor.

The Kerr-like black hole solution to the field equation~\eqref{fieldequation} was derived in~\cite{Ding2019,Liu2019}
{\small\begin{eqnarray}
\label{Kerrlm}
\nn ds^2&=&-\left(1-\frac{2M r}{\rho^2}\right)dt^2-\frac{4M a r \lambda\sin^2 \theta}{\rho^2}d\varphi dt\\
&&+\frac{\rho^2}{\Delta}dr^2+\rho^2d\theta^2+\frac{A\sin^2 \theta}{\rho^2}d\varphi^2~,~~~
\end{eqnarray}}
where
{\begin{eqnarray}
\label{Kerrlm}
&&\nn\lambda=\sqrt{1+l}~,\\
&&\nn \rho^2=r^2+\lambda^2a^2\cos^2\theta~.\\
&&\nn \Delta=\frac{r^2-2Mr}{\lambda^2}+a^2~,\\
&&\nn A=[r^2+\lambda^2a^2]^2-\Delta\lambda^4a^2\sin^2\theta~.
\end{eqnarray}}
In the above, $M$ and $a$ are the mass and rotating angular
momentum of the black hole respectively, $l$ is Lorentz violation parameter. For $l=0$, the Kerr-like solution leads to the Kerr solution in GR~\cite{BL1967}. In addition, the Schwarzschild-like solution in bumblebee
gravity model is covered as $a=0$~\cite{Casana2018}.

The event horizons and ergosphere of black hole locate at
{\begin{eqnarray}
\label{Kerrlm}
&&\nn r_{\pm}=M\pm\sqrt{M^2-a^2\left(1+l\right)}~,\\
&&\nn r_{\pm}^{ergo}=M\pm\sqrt{M^2-a^2\left(1+l\right)\cos^2\theta}~,
\end{eqnarray}}
where the positive and negative sign is for outer and inner horizon / ergosphere, respectively. The existence of black holes requires~\cite{Ding2019,Liu2019}
{\begin{eqnarray}
&&\nn \vert a \vert\leq \frac{M}{\lambda}~.
\end{eqnarray}}
\subsection{Motion of massive particles on the equatorial plane}
 The relativistic Lagrangian for a free particle moving on the equatorial plane $(\theta=\pi/2)$ of the Kerr-like black hole is
 \begin{eqnarray}
\label{Lagrangian}
 2\mathcal{L}&=&-m g_{\mu \nu} \dot{x}^{\mu} \dot{x}^{\nu}\nn\\
 &=&m\left(1-\frac{2M}{r}\right)\dot{t}^2+m\frac{4M a \lambda}{r}\dot{t}\dot{\varphi}\nn\\
&&-m\frac{r^2}{\Delta}\dot{r}^2-m\frac{A}{r^2}\dot{\varphi}^2~,
\end{eqnarray}
where a dot denotes the the differentiation with respect to an arbitrary parameter. Then one can obtain two conserved quantities as following
\begin{eqnarray}
\label{canonical momentum1}
&& p_0=\frac{\partial \mathcal{L}}{\partial \dot{t}}=m\left(1-\frac{2M}{r}\right)\dot{t}+m\frac{2M a \lambda}{r}\dot{\varphi}=\mathcal{E}~,~~~\\
\label{canonical momentum2}
&& p_\varphi=\frac{\partial \mathcal{L}}{\partial \dot{\varphi}}=m\frac{2M a \lambda}{r}\dot{t}-m\frac{A}{r^2}\dot{\varphi}=-\mathcal{J}~,
\end{eqnarray}
where $\mathcal{E}$ and $\mathcal{J}$ are the conserved energy and angular momentum of the particle, respectively. They can be measured at infinity for an asymptotic observer by
\begin{eqnarray}
\label{EJ}
&& \mathcal{E}=\frac{m}{\sqrt{1-v^2}}~,~\mathcal{J}=\frac{mvb}{\sqrt{1-v^2}}~,
\end{eqnarray}
where $v$ is the particle velocity and $b$ is the impact parameter defined by
\begin{eqnarray}
\label{impact parameter}
&& bv\equiv\frac{\mathcal{J}}{\mathcal{E}}~.
\end{eqnarray}
For convenience, we can choose an appropriate parameter so that $2\mathcal{L}=m$. Then, using Eqs.~\eqref{Lagrangian}-~\eqref{EJ}, the orbit equation of massive particle can be obtained by following
\begin{eqnarray}
\label{Sorbit}
\left(\frac{du}{d\varphi}\right)^2&=&\frac{1}{\lambda^2}\left(\frac{1}{b^2}-u^2\right)+\frac{2M u\left(1-v^2+b^2u^2v^2\right)}{b^2v^2\lambda^2}\nn\\
&&-\frac{4M a u}{b^3 v \lambda}+\mathcal{O}\left(M^2,a^2\right)~.~~~~
\end{eqnarray}
where $u=1/r$ is the inverse radial coordinate.
Considering the condition $\frac{du}{d\varphi}\mid_{\varphi=\frac{\lambda \pi}{2}}=0$, the above equation can be solved iteratively as
\begin{eqnarray}
\label{orbit}
u(\varphi)&=&\frac{\sin\left(\frac{\varphi}{\lambda}\right)}{b}+\frac{1+v^2\cos^2\left(\frac{\varphi}{\lambda}\right)}{b^2v^2}M\nn\\
&&-\frac{2M a\lambda}{b^3 v}+\mathcal{O}\left(M^2,a^2\right)~.
\end{eqnarray}
In addition, we can obtain the iterative solution for $\varphi$ in the above equation as
{\begin{eqnarray}
\label{CTA}
&&\varphi=\begin{cases}
\varphi_1-M\varphi_2+aM\varphi_3+...~,    &\text{if } \vert{\varphi}\vert <\frac{\lambda\pi}{2}~;\\
\lambda\pi-\varphi_1+M\varphi_2-aM\varphi_3+...~,&\text{if } \vert {\varphi}\vert >\frac{\lambda\pi}{2}~,~~~~
\end{cases}\
\end{eqnarray}}
where
{\begin{eqnarray}
&&\varphi_1=\lambda\arcsin{(bu)}~,\nn\\
&&\varphi_2=\frac{\left(1+v^2-b^2u^2v^2\right)\lambda}{b^2\sqrt{1-b^2u^2}v^2}~,\nn\\
&&\varphi_3=\frac{2\lambda^2}{b^2\sqrt{1-b^2u^2}v}~.\nn
\end{eqnarray}}

\section{The gravitational deflection angle using the Gauss-Bonnet theorem}\label{GBDEF}
\subsection{Jacobi-Maupertuis Randers-Finsler metric}
As mentioned in the introduction, we need the Jacobi metric of curved spacetime in order to use the GB theorem to study the deflection angle of massive particles. In particular, the Jacobi metric of stationary spacetime corresponds to the JMRF metric. The trajectories of neutral particles moving in a stationary are seen as the geodesics of the corresponding JMRF metric space. The JMRF metric in our spacetime signature can be written~\cite{Chanda2019}
\begin{eqnarray}
\label{Randers-Finsler}
ds_J&=&\sqrt{\frac{\mathcal{E}^2+m^2 g_{00}}{-g_{00}}\gamma_{ij}dx^i dx^j}-\mathcal{E} \frac{g_{0i}}{g_{00}}dx^i\nn\\
&\equiv& \sqrt{\alpha_{ij}dx^i dx^j}+\beta_i dx^i~,
\end{eqnarray}
where the spatial metric $\gamma_{ij}$ is defined by
\begin{eqnarray}
\label{spatial metric}
&& \gamma_{ij}:=g_{ij}-\frac{g_{0i} g_{0j}}{g_{00}}~.~~~
\end{eqnarray}
In Eq.~\eqref{Randers-Finsler}, $\alpha_{ij}$ is a Riemannian metric and $\beta_i$ is a one-form, which satisfies the positivity and convexity~\cite{Chern2002}
\begin{eqnarray}
&&\sqrt{\alpha^{ij}\beta_i \beta_j}<1~,
\end{eqnarray}
From the Kerr-like metric~\eqref{Kerrlm}, one has
\begin{eqnarray}
g_{00}&=&-\left(1-\frac{2M r}{\rho^2}\right)~,~g_{0\varphi}=-\frac{2M a r \lambda\sin^2 \theta}{\rho^2}~,\nn\\
\gamma_{ij}dx^i dx^j&=&\frac{\rho^2}{\Delta}dr^2+\rho^2d\theta^2\nn\\
&&+\left[\frac{A\sin^2\theta}{\rho^2}+\frac{4a^2\lambda^2M^2r^2\sin^4\theta}{(1-\frac{2Mr}{\rho^2})\rho^4}\right]d\varphi^2.\nn\\
\end{eqnarray}
Then by Eq.~\eqref{Randers-Finsler}, we find the following JMFR metric
{\begin{eqnarray}
\label{KerrJMRE}
\alpha_{ij}dx^i dx^j&=&\left(\frac{{\mathcal{E}}^2}{1-\frac{2Mr}{\rho^2}}-m^2\right)\bigg[\frac{\rho^2dr^2}{\Delta}+\rho^2d\theta^2\nn\\
&&+\left(\frac{A\sin^2\theta}{\rho^2}+\frac{4a^2\lambda^2M^2r^2\sin^4\theta}{(1-\frac{2Mr}{\rho^2})\rho^4}\right)d\varphi^2\bigg]~,\nn\\
\beta_i dx^i&=&-\frac{2\lambda\mathcal{E} M a r \sin^2\theta}{r(r-2M)+a^2\lambda^2\cos^2\theta}d\varphi~.
\end{eqnarray}}
Note that this equation can leads to the optical Finsler-Randers metric when $\mathcal{E}=1$ and $m=0$.

\subsection{The generalized Jacobi metric method}
By replacing optical metric with Jacobi metric , one can extend the generalized optical method to calculate the deflection of massive particles~\cite{LJ2019}. We call the positive Riemannian metric $\alpha_{ij}$ as generalized Jacobi metric and suppose that the particles live in the Remannian space $\bar{M}$ defined by the generalized Jacobi metric, that is
\begin{eqnarray}
\label{GJM}
&& d\sigma^2=\alpha_{ij}dx^idx^j~.
\end{eqnarray}
 By the way, the particle ray now is not the geodesic in $\bar{M}$ and its geodesic curvature can be calculated by~\cite{OIA2017}
\begin{eqnarray}
\label{gc2}
&& k_g=-\frac{\beta_{\varphi,r}}{\sqrt{{det{\alpha}~\alpha^{\theta\theta}}}}~,
\end{eqnarray}
where the comma denotes the partial derivative.

The deflection angle can be defined by~\cite{OIA2017}
\begin{eqnarray}
\label{angle}
&& \hat{\alpha}\equiv \Psi_R-\Psi_S+\varphi_{RS}~,
\end{eqnarray}
 where $\Psi_R$ and $\Psi_S$ are angles between the particle ray tangent and the radial direction from the lens in receiver and source, respectively, and the coordinate angle $\varphi_{RS}\equiv\varphi_R-\varphi_S$.

Next, we shall use GB theorem to study the gravitational lensing. The GB theorem reveals the relation between geometry and topology of surface. Suppose that $D$ is a subset of a compact, oriented surface, with Gaussian curvature $K$ and Euler characteristic $\chi(D)$. Its boundary $\partial{D}$ is a piecewise smooth curve with geodesic curvature $k$. In the i-th vertex for $\partial{D}$, the jump angle denotes $\phi_i$, in the positive sense. Then, the GB theorem states that~\cite{GW2008,Carmo1976}:
\begin{equation}
\iint_D{K}dS+\oint_{\partial{D}}k~d\sigma+\sum_{i=1}{\phi_i}=2\pi\chi(D)~,\\
\end{equation}
where $dS$ is the area element of the surface and $d\sigma$ is the line element along the boundary.

\begin{figure}[t]
\centering
\includegraphics[width=8.5cm]{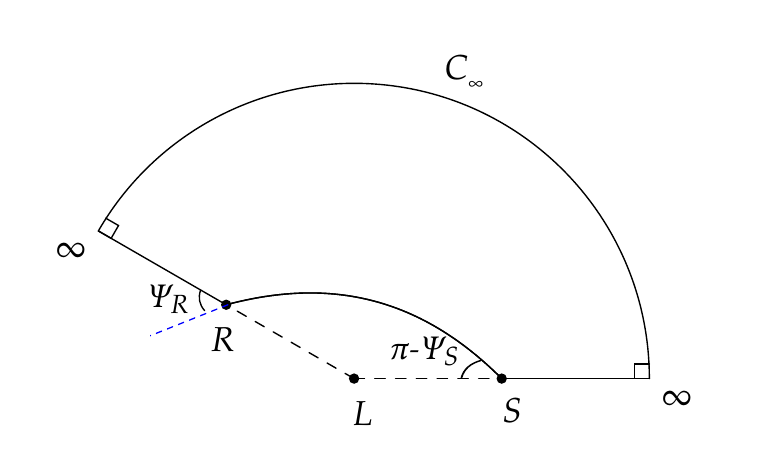}
\caption{The quadrilateral $\prescript{\infty}{R}\Box_{S}^{\infty}\subset(\bar{M},\alpha_{ij})$. R, S and L denote the receiver, the source and the lens, respectively. $\Psi_R$ and $\Psi_S$ are angles between the particle ray tangent and the radial direction from the lens in R and S, respectively. The curve $C_{\infty}$ denotes $C_{r_0}(r_0\rightarrow \infty)$ and $C_{r_0}$ is defined by $r(\varphi)=r_0=constant$. Note that each outer angle at the intersection of the radial direction curves and $C_{\infty}$ is $\pi/2$.}\label{Figure}
\end{figure}

Then, we consider the quadrilateral $\prescript{\infty}{R}\Box_{S}^{\infty}\subset(\bar{M},\alpha_{ij})$, as shown in Fig.~\ref{Figure}. It is bounded by four curves: the trajectory of particle connection source (S) and receiver (R), two spatial geodesics of outgoing radial lines from R and S respectively, and a circular arc segment $C_{\infty}$, where $C_{\infty}$ denotes $C_{r_0}(r_0\rightarrow \infty)$ and $C_{r_0}$ is defined by $r(\varphi)=r_0=constant$. For curve $C_{\infty}$, we have $kdl=\frac{1}{\lambda} d\varphi$ and thus $\int_{C_{\infty}}k dl=\frac{1}{\lambda}\varphi_{RS}$. In addition, the Euler characteristic of this quadrilateral is unity. Notice that the sum of two jump angles in infinite is $\pi$. In addition, we have $\phi_S=\pi-\Psi_S$ and $\phi_R=\Psi_R$. Finally, using GB theorem to the quadrilateral leads to

 {\small\begin{equation}
\int\int_{_{R}^{\infty}\Box_{S}^{\infty}}K dS-\int_{S}^{R}k_gdl+\frac{1}{\lambda}\varphi_{RS}+\Psi_R-\Psi_S=0~.\\
\end{equation}}
 By this expression, Eq.~\eqref{angle} can be rewritten as
{\begin{eqnarray}
\label{gbdef}
\hat{\alpha}&=&-\int\int_{_{R}^{\infty}\Box_{S}^{\infty}}K dS+\int_{S}^{R}k_g d\sigma+\left(1-\frac{1}{\lambda}\right)\varphi_{RS}~.\nn\\
\end{eqnarray}}
This expression is different with the asymptotically
flat case~\cite{LJ2019} due to the existence of the third part to the right of the equal sign. It obvious that the term $\varphi_{RS}$ vanishes if $\lambda=1$ ($l=0$). In the above, the Gaussian curvature of generalized Jacobi metric can be calculated by~\cite{Werner2012}.
{\small\begin{eqnarray}
\label{Gauss-K}
{K}&=&\frac{\bar{R}_{r \varphi r \varphi}}{det{\alpha}}\nn\\
&=&\frac{1}{\sqrt{det \alpha}}\left[\frac{\partial}{\partial{\varphi}}\left(\frac{\sqrt{det \alpha}}{\alpha_{rr}}{{\bar{\Gamma}}^\varphi_{rr}}\right)-\frac{\partial}{\partial{r}}\left(\frac{\sqrt{det \alpha}}{\alpha_{rr}}{{\bar{\Gamma}}^\varphi_{r\varphi}}\right)\right]~,~~~~~
\end{eqnarray}}
where the quantities associated with Jacobi metric $\alpha_{ij}$ are added a bar above. Notice that in Eq.~\eqref{Gauss-K} we only need the two-dimensional generalized Jacobi metric due to this paper only considering the equatorial plane case.

\subsection{The gravitational deflection angle of massive particles}
In this subsection, we shall apply Eq.~\eqref{gbdef} to compute the deflection angle.
\subsubsection{$\varphi_{RS}$ part}
We first consider the part of $\varphi_{RS}$. By the lensing setup, Eq.~\eqref{CTA} leads to
{\begin{eqnarray}
\label{SKerr}
\varphi_S&=&\lambda\arcsin{(bu_S)}-\frac{\left(1+v^2-b^2u_S^2v^2\right)\lambda}{b\sqrt{1-b^2u_S^2}v^2}M\nn\\
&&+\frac{2M a \lambda^2}{b^2\sqrt{1-b^2u_S^2}v}+\mathcal{O}(M^2,a^2)~,\\
\label{RKerr}
\varphi_R&=&\lambda\left[\pi-\arcsin{(bu_R)}\right]+\frac{\left(1+v^2-b^2u_R^2v^2\right)\lambda}{b\sqrt{1-b^2u_R^2}v^2}M\nn\\
&&-\frac{2M a \lambda^2}{b^2\sqrt{1-b^2u_R^2}v}+\mathcal{O}(M^2,a^2)~.
\end{eqnarray}}
It is convenient to use $u_R$ and $u_S$ to express the finite-distance deflection angle. Then, we can get
{\begin{eqnarray}
\label{RSDEF}
\nn\varphi_{RS}&=&\varphi_R-\varphi_S\nn\\
&=&\pi \lambda-\lambda\left[\arcsin(b u_R)+\arcsin(b u_S)\right]\nn\\
\nn&&+\frac{M \lambda}{bv^2}\left(\frac{1}{\sqrt{1-b^2u_R^2}}+ \frac{1}{\sqrt{1-b^2u_S^2}}\right)\\
\nn&&+\frac{M \lambda}{b}\left(\sqrt{1-b^2u_R^2}+\sqrt{1-b^2u_S^2}\right)\\
&&-\frac{2a M\lambda^2}{b^2v}\left(\frac{1}{\sqrt{1-b^2u_R^2}}+\frac{1}{\sqrt{1-b^2u_S^2}}\right)~.~~~
\end{eqnarray}}
\subsubsection{Gaussian curvature}
Considering Eq.~\eqref{KerrJMRE}, we can find the generalized Kerr-like Jacobi metric in the equatorial plane $(\theta=\pi/2)$ as following
{\begin{eqnarray}
\label{GKJM}
d\sigma^2&=&\alpha_{ij}dx^i dx^j\nn\\
&=&\mathcal{E}^2\left(\frac{1}{1-\frac{2M}{r}}-1+v^2\right)\bigg[\frac{\lambda^2r^2dr^2}{a^2\lambda^2+r\left(r-2M\right)}\nn\\
&&+\frac{r\left(a^2\lambda^2+r^2-2Mr\right)}{r-2M}d\varphi^2\bigg]~,
\end{eqnarray}}
where we use Eq.~\eqref{EJ}. Bring the corresponding metric components in Eq.~\eqref{GKJM} into Eq.~\eqref{Gauss-K}, we can obtain the Gaussian curvature up to leading order as following
{\begin{eqnarray} K&=&-\frac{\left(1+v^2\right)M}{\mathcal{E}^2r^3v^4\lambda^2}+\mathcal{O}(M^2,a^2)~.
\end{eqnarray}}
Then the surface integral of Gaussian curvature is given by
{\begin{eqnarray}
\label{GaussDEF}
&&-\int\int_{_{R}^{\infty}\Box_{S}^{\infty}}K dS\nn\\
&=&\int_{\varphi_S}^{\varphi_R} \int_{r(\varphi)}^{\infty} \left[\frac{\left(1+v^2\right)M}{r^2 v^2 \lambda}+\mathcal{O}\left(M^2,a^2\right)\right]dr d\varphi~\nn\\
&=&\frac{M\left(\sqrt{1-b^2u_R^2}+\sqrt{1-b^2u_S^2}\right)\left(1+v^2\right)}{b v^2}\nn\\
&&+\mathcal{O}\left(M^2,a^2\right)~,
\end{eqnarray}}
where we have used Eq.~\eqref{orbit}, Eq.~\eqref{SKerr} and Eq.~\eqref{RKerr}. Note that the integration of the Gaussian curvature is independent of $\lambda$.
\subsubsection{Geodesic curvature}
Now we calculate the geodesic curvature of particle ray. Substituting corresponding quantities in Eq.~\eqref{KerrJMRE} into Eq.~\eqref{gc2}, the geodesic curvature of particles ray can be obtained as
{\begin{eqnarray}
\label{GEKerr}
 &&k_g=-\frac{2a M }{\mathcal{E}r^3v^2}+\mathcal{O}(M^2,a^2)~,~~~~
\end{eqnarray}}
where we use Eq.~\eqref{EJ}. In addition, Eq.~\eqref{GKJM} deduces the 0th-order parameter transformation
{\begin{eqnarray}
\label{PTKerr}
 d\sigma=\mathcal{E} b v\csc^2\left(\frac{\varphi}{\lambda}\right)~ d\varphi+\mathcal{O}\left(M,a\right)~,~~~~
\end{eqnarray}}
where we have used $r=b/\sin\left(\frac{\varphi}{\lambda}\right)$.
Then, one can get the part of deflection angle related with the path integral of geodesic curvature of the particle ray. Considering Eqs.~\eqref{GEKerr} and~\eqref{PTKerr}, we have
{\begin{eqnarray}
\label{GeodesicDEF}
 \int_{S}^{R}k_gd\sigma&=&-\frac{2a M }{b^2v}\int_{\varphi_S}^{\varphi_R}\sin\left(\frac{\varphi}{\lambda}\right)d\varphi+\mathcal{O}(M^2,a^2)\nn\\
 &=&-\frac{2a M\lambda  \left(\sqrt{1-b^2u_R^2}+\sqrt{1-b^2u_S^2}\right)}{b^2v}\nn\\
 &&+\mathcal{O}(M^2,a^2)~,~~~~~~~~
\end{eqnarray}}
where we use Eq.~\eqref{orbit}, Eq.~\eqref{SKerr} and Eq.~\eqref{RKerr}.

\subsubsection{Deflection angle}
Considering Eq.~\eqref{GaussDEF}, Eq.~\eqref{RSDEF} and Eq.~\eqref{GeodesicDEF}, the finite-distance deflection angle of massive particles in Kerr-like spacetime can be obtained as following
{\begin{eqnarray}
\label{Kerr-like}
 \hat{\alpha}&=&-\int\int_{_{R}^{\infty}\Box_{S}^{\infty}}K dS+\int_{S}^{R}k_g d\sigma+\left(1-\frac{1}{\lambda}\right)\varphi_{RS}\nn\\
 &=&\left(\lambda-1\right)\left[\pi-\arcsin(b u_R)-\arcsin(b u_S)\right]\nn\\
 &&+\bigg[\frac{\left(1+v^2\right)\lambda-b^2 u_R^2\left(1+v^2\lambda\right)}{\sqrt{1-b^2u_R^2}}\nn\\
 &&+\frac{\left(1+v^2\right)\lambda-b^2 u_S^2\left(1+v^2\lambda\right)}{\sqrt{1-b^2u_S^2}}\bigg]\frac{M}{bv^2}\nn\\
 &&-\frac{2a M  \lambda}{b^2 v}\left(\frac{\lambda-b^2 u_R^2}{\sqrt{1-b^2u_R^2}}+\frac{\lambda-b^2 u_S^2}{\sqrt{1-b^2u_S^2}}\right)\nn\\
 &&+\mathcal{O}(M^2,a^2)~.
\end{eqnarray}}

\section{The gravitational deflection angle by definition}\label{DEFDEF}
In the previous section, we calculated the deflection angle by GB theorem from the perspective of geometry and topology. This section will directly calculate the deflection angle by the definition in Eq.~\eqref{angle}
\begin{eqnarray}
\label{angle1}
&& \hat{\alpha}\equiv \Psi_R-\Psi_S+\varphi_{RS}~.
\end{eqnarray}
Notice that $\varphi_{RS}$ has been obtained in Eq.~\eqref{RSDEF}. Now we calculate the angles $\Psi_R$ and $\Psi_S$. We still suppose that the particles lives in the generalized Jacobi metric space $(\bar{M}, \alpha_{ij})$. Then the unit tangent vector of particle ray in the equatorial plane can be written as
\begin{eqnarray}
&& e^i=\frac{d x^i}{d\sigma}=\left(\frac{dr}{d\sigma},0,\frac{d\varphi}{d\sigma}\right)~.
\end{eqnarray}
Choosing the outgoing direction, the unit radial vector in the equatorial plane can be written as
\begin{eqnarray}
&& R^i=\left(\frac{1}{\sqrt{\alpha_{rr}}},0,0\right)~.
\end{eqnarray}
Then, we have
\begin{eqnarray}
\cos\Psi&=&\alpha_{ij}e^i R^j\nn\\
&=&\sqrt{\alpha_{rr}}(\frac{dr}{d\sigma})\nn\\
&=&\frac{\sqrt{\alpha_{rr}}}{\sqrt{\alpha_{rr}+\alpha_{\varphi\varphi}(\frac{d\varphi}{dr})^2}}~,
\end{eqnarray}
where we used $\alpha_{ij}e^i e^j=1$ and $\alpha_{ij}R^i R^j=1$.
This can be rewritten as
 \begin{eqnarray}
 \label{ang}
\sin\Psi&=&\frac{\sqrt{\alpha_{\varphi\varphi}}}{\sqrt{\alpha_{rr}(\frac{dr}{d\varphi})^2+\alpha_{\varphi\varphi}}}~.
\end{eqnarray}
Bring the corresponding metric components in Eq.~\eqref{GKJM} into Eq.~\eqref{ang} and using Eq.~\eqref{orbit}, we finally arrive at
\begin{eqnarray}
\sin\Psi&=&\sin\left(\frac{\varphi}{\lambda}\right)+\frac{M\cos\left(\frac{\varphi}{\lambda}\right)}{bv^2}\bigg[(1+v^2)\cos^2\left(\frac{\varphi}{\lambda}\right)\nn\\
&&-2v^2\cos\left(\frac{\pi}{2\lambda}\right)\bigg]-\frac{2a M \lambda \cos^2\left(\frac{\varphi}{\lambda}\right)}{b^2 v}\nn\\
&&+\mathcal{O}\left(M^2,a^2\right)~.
\end{eqnarray}
Then, we obtain the following results
{\begin{eqnarray}
\label{0R-ABC}
\Psi_S&=&\pi-\arcsin(b u_S)+\frac{M}{bv^2}\bigg[2v^2\cos\left(\frac{\pi}{2\lambda}\right)+\frac{b^2u_S^2}{\sqrt{1-b^2u_S^2}}\bigg]\nn\\
&&-\frac{2a M u_S^2\lambda}{\sqrt{1-b^2u_S^2}v}+\mathcal{O}\left(M^2,a^2\right)~,\\
\Psi_R&=&\arcsin(b u_R)+\frac{M}{bv^2}\bigg[2v^2\cos\left(\frac{\pi}{2\lambda}\right)-\frac{b^2u_R^2}{\sqrt{1-b^2u_R^2}}\bigg]\nn\\
&&+\frac{2a M u_R^2\lambda}{\sqrt{1-b^2u_R^2}v}+\mathcal{O}\left(M^2,a^2\right)~.
\end{eqnarray}}
Thus, the gravitational angle is
{\begin{eqnarray}
\hat{\alpha}&=& \Psi_R-\Psi_S+\varphi_{RS}\nn\\
&=&\left(\lambda-1\right)\left[\pi-\arcsin(b u_R)-\arcsin(b u_S)\right]\nn\\
 &&+\bigg[\frac{\left(1+v^2\right)\lambda-b^2 u_R^2\left(1+v^2\lambda\right)}{\sqrt{1-b^2u_R^2}}\nn\\
 &&+\frac{\left(1+v^2\right)\lambda-b^2 u_S^2\left(1+v^2\lambda\right)}{\sqrt{1-b^2u_S^2}}\bigg]\frac{M}{bv^2}\nn\\
 &&-\frac{2a M  \lambda}{b^2 v}\left(\frac{\lambda-b^2 u_R^2}{\sqrt{1-b^2u_R^2}}+\frac{\lambda-b^2 u_S^2}{\sqrt{1-b^2u_S^2}}\right)\nn\\
 &&+\mathcal{O}(M^2,a^2)~.
\end{eqnarray}}
This result is exactly the same as Eq.~\eqref{Kerr-like}.
\section{Finite-distance corrections}\label{FDC}
In our calculation, the particle orbits are assumed as prograde relative to the rotation of the Kerr-like black hole. The sign of the term including angular momentum $a$ changes if the particles ray is a retrograde orbit. In short, the deflection angle can be written as
{\begin{eqnarray}
\label{Kerrlike}
 \hat{\alpha}&=&\left(\lambda-1\right)\left[\pi-\arcsin(b u_R)-\arcsin(b u_S)\right]\nn\\
 &&+\bigg[\frac{\left(1+v^2\right)\lambda-b^2 u_R^2\left(1+v^2\lambda\right)}{\sqrt{1-b^2u_R^2}}\nn\\
 &&+\frac{\left(1+v^2\right)\lambda-b^2 u_S^2\left(1+v^2\lambda\right)}{\sqrt{1-b^2u_S^2}}\bigg]\frac{M}{bv^2}\nn\\
 &&\pm\frac{2a M  \lambda}{b^2 v}\left(\frac{\lambda-b^2 u_R^2}{\sqrt{1-b^2u_R^2}}+\frac{\lambda-b^2 u_S^2}{\sqrt{1-b^2u_S^2}}\right)\nn\\
 &&+\mathcal{O}(M^2,a^2)~,
\end{eqnarray}}
where $\pm$ signs correspond to retrograde and prograde particle orbits, respectively. For $\lambda=1$ ($l=0$), Eq.~\eqref{Kerrlike} agree with the result in the Kerr spacetime~\cite{LJ2019}. 
\begin{figure}[t]
\begin{center}
\includegraphics[width=0.4\textwidth]{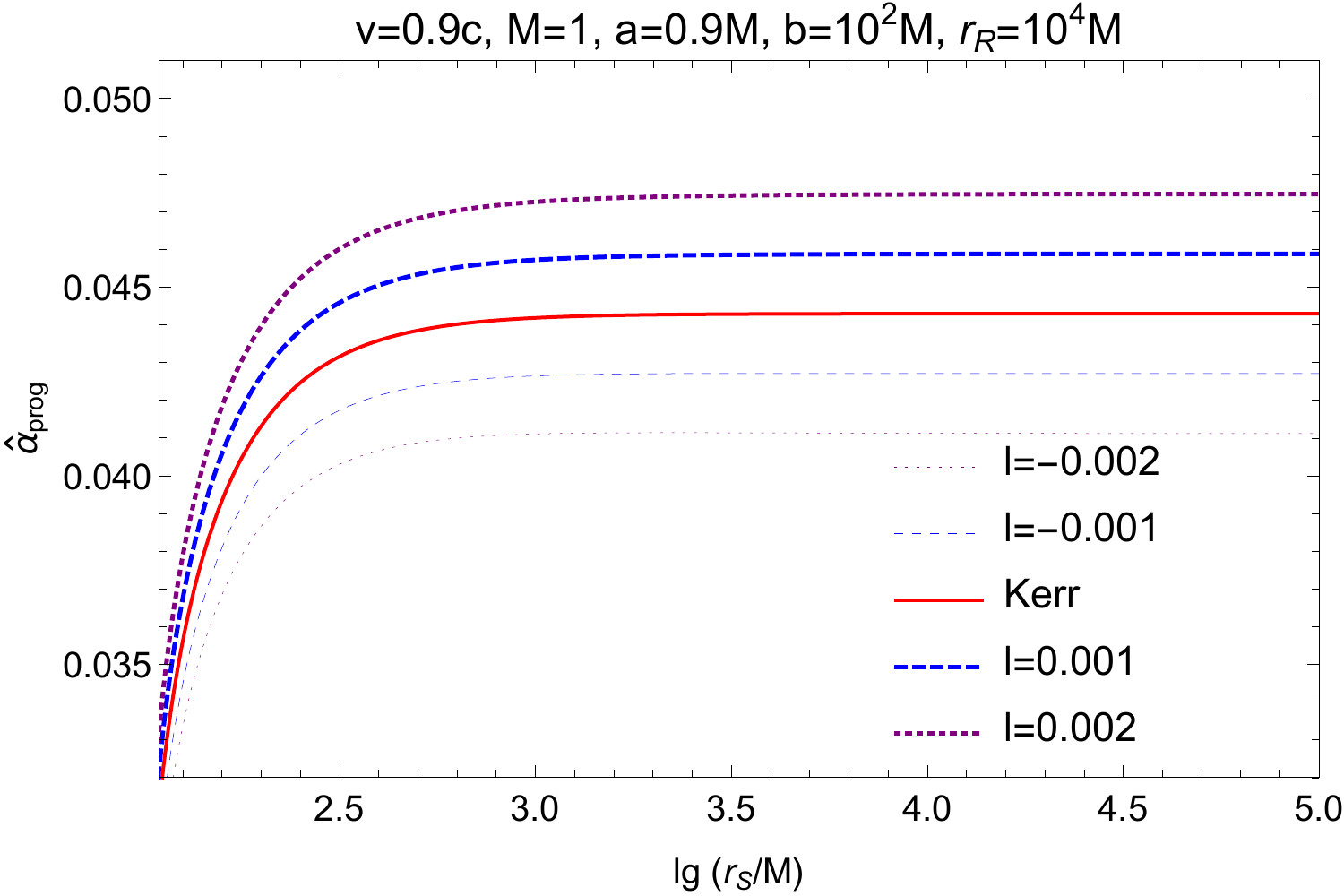}\hspace{0.04\textwidth}\\
(a)\hspace{0.3\textwidth}\\
\includegraphics[width=0.4\textwidth]{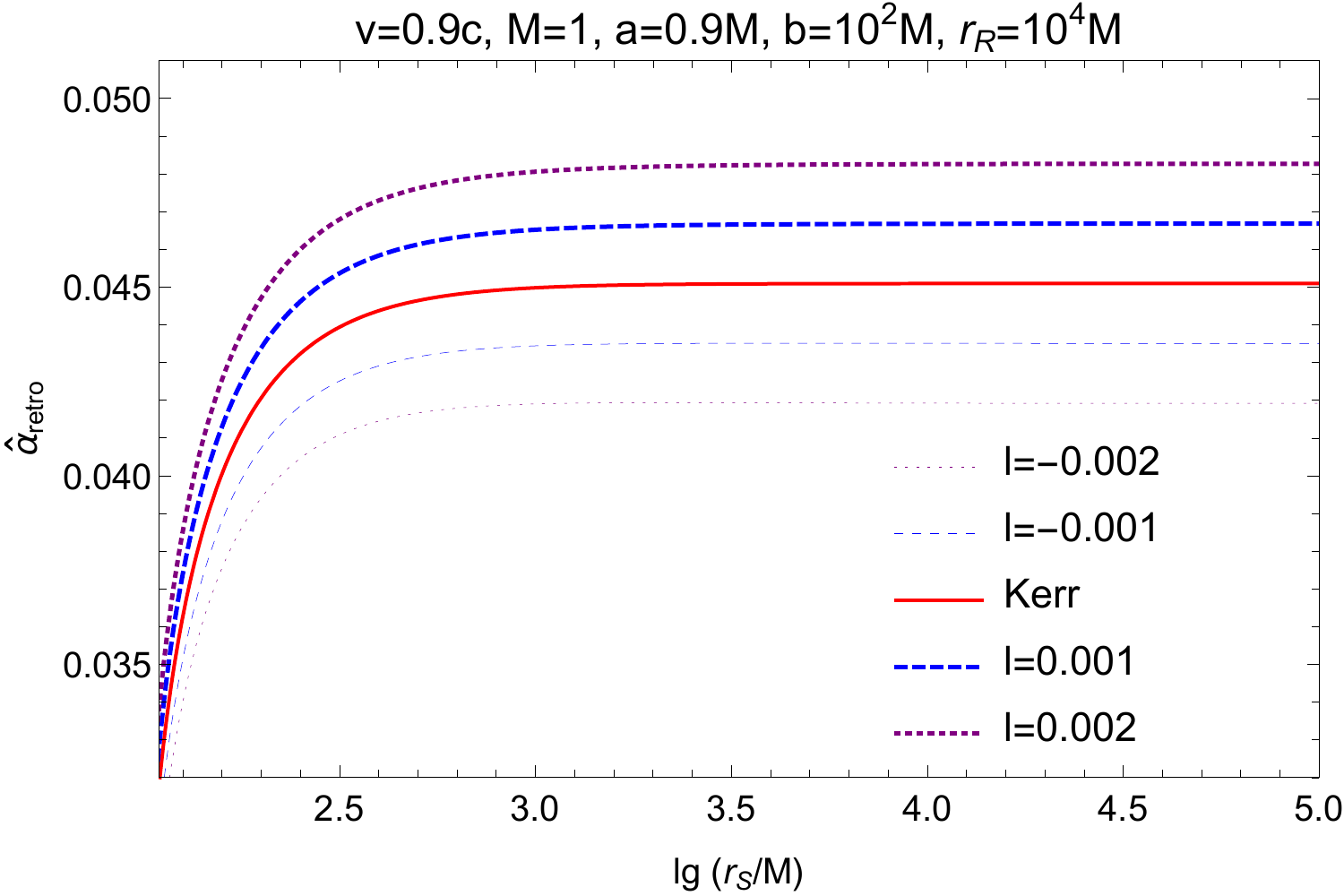}\hspace{0.04\textwidth}\\
(b)\hspace{0.3\textwidth}\\
\caption{The finite-distance deflection angle $\hat{\alpha}$. The vertical axis denotes $\hat{\alpha}$, and the horizontal axis denotes $\lg~(r_S/M)$. We suppose $v=0.9c$, $M=1$, $a=0.9M$, $b=10^2M$, and $r_R=10^4M$. The red solid line, thin purple dotted line, thin blue dashed line, thick blue dashed line, thick purple dotted line correspond to $l=0$ (Kerr spacetime), $l=-0.002$, $l=-0.001$, $l=0.001$ and $l=0.002$, respectively. The deflection angle of prograde particle orbit corresponds to (a) and retrograde particle orbit corresponds to (b).} \label{figangle}
\end{center}
\end{figure}

In Fig.~\ref{figangle}, we plot finite-distance deflection angle $\hat{\alpha}$ against $\lg (r_S/M)$. Here and henceforth we have supposed $v=0.9c$, $M=1$, $a=0.9M$, $b=10^2M$, and $r_R=10^4M$. (a) and (b) are the deflection angle of prograde and retrograde particle orbits, respectively. We chose Lorentz breaking constant $l=0$, $l=\pm0.001$ and $l=\pm0.002$ to picture five lines. From which, we can see that the deflection angle monotonically increases as $r_S$ increases. The deflection angle also increases as $l$ increases. Furthermore, the deflection angle is larger than it in Kerr spacetime as $l>0$, while it is smaller as $l<0$.

The deflection angle up to second-order in Lorentz breaking constant $l$ reads
{\begin{eqnarray}
\hat{\alpha}&=&\left(\frac{1}{2}-\frac{l}{8}\right)l\left[\pi-\arcsin(b u_R)-\arcsin(b u_S)\right]\nn\\
 &&+\frac{M\left(\sqrt{1-b^2u_R^2}+\sqrt{1-b^2u_S^2}\right)\left(1+v^2\right)}{bv^2}\nn\\
 &&\pm\frac{2M a\left(\sqrt{1-b^2u_R^2}+\sqrt{1-b^2u_S^2}\right)}{b^2 v}\nn\\
 &&+\frac{\left(4-l\right)M l}{8bv^2}\times\nn\\
 &&\left[\frac{1+\left(1-b^2u_R^2\right)v^2 }{\sqrt{1-b^2u_R^2}}+\frac{1+\left(1-b^2u_S^2\right)v^2 }{\sqrt{1-b^2u_S^2}}\right]\nn\\
 &&\pm\frac{a M  l}{b^2 v}\left(\frac{2-b^2 u_R^2}{\sqrt{1-b^2u_R^2}}+\frac{2-b^2 u_S^2}{\sqrt{1-b^2u_S^2}}\right)\nn\\
 &&\pm\frac{a M  l^2}{4v}\left(\frac{u_R^2}{\sqrt{1-b^2u_R^2}}+\frac{ u_S^2}{\sqrt{1-b^2u_S^2}}\right)\nn\\
 &&+\mathcal{O}(M^2,a^2,l^3)~.
\end{eqnarray}}
Obviously, these terms excluding angular momentum $a$ are positive if they are at the order in $l$,  whereas they are negative if they at the order in $l^2$. Interestingly, the terms containing $a$ are both positive (retrograde orbits) or negative (prograde orbits), regardless of whether they are at the order in $l$ or the order in $l^2$. In general, the deflection angle increases as $l$ increases, as shown in Fig.~\ref{figangle}.

Now we consider several limits. First, Eq.~\eqref{Kerrlike} leads to the finite-distance deflection of light ($v=1$),
{\begin{eqnarray}
\label{light-finite}
 \hat{\alpha}_{v=1}&=&\left(\lambda-1\right)\left[\pi-\arcsin(b u_R)-\arcsin(b u_S)\right]\nn\\
 &&+\bigg[\frac{2\lambda-b^2 u_R^2\left(1+\lambda\right)}{\sqrt{1-b^2u_R^2}}+\frac{2\lambda-b^2 u_S^2\left(1+\lambda\right)}{\sqrt{1-b^2u_S^2}}\bigg]\frac{M}{b}\nn\\
 &&\pm\frac{2a M  \lambda}{b^2}\left(\frac{b^2 u_R^2-\lambda}{\sqrt{1-b^2u_R^2}}+\frac{b^2 u_S^2-\lambda}{\sqrt{1-b^2u_S^2}}\right)\nn\\
 &&+\mathcal{O}(M^2,a^2)~.
\end{eqnarray}}
Secondly, for $u_R\rightarrow0$ and $u_S\rightarrow0$, Eq.~\eqref{Kerrlike} leads to the infinite-distance deflection of massive particles,
{\begin{eqnarray}
\label{massive-infinite}
 \hat{\alpha}_{\infty}&=&\left(\lambda-1\right)\pi+\frac{2M \left(1+v^2\right)\lambda}{bv^2}\pm\frac{4a M \lambda^2}{b^2v}\nn\\
 &&+\mathcal{O}(M^2,a^2)~,
\end{eqnarray}}
which agree with the result obtained by Werner's Finsler geometry method, given in Appendix \ref{AppendixA}.
Finally, For $a=0$, Eq.~\eqref{Kerrlike} leads to the finite-distance deflection of massive particles by Schwarzschild-like solution in the bumblebee
gravity model,
{\begin{eqnarray}
\label{Schwarzschild-like}
 \hat{\alpha}_S&=&\left(\lambda-1\right)\left[\pi-\arcsin(b u_R)-\arcsin(b u_S)\right]\nn\\
 &&+\bigg[\frac{\left(1+v^2\right)\lambda-b^2 u_R^2\left(1+v^2\lambda\right)}{\sqrt{1-b^2u_R^2}}\nn\\
 &&+\frac{\left(1+v^2\right)\lambda-b^2 u_S^2\left(1+v^2\lambda\right)}{\sqrt{1-b^2u_S^2}}\bigg]\frac{M}{bv^2}\nn\\
 &&+\mathcal{O}(M^2)~.
\end{eqnarray}}
It is worth pointing out that the Eq.~\eqref{light-finite}, Eq.~\eqref{massive-infinite} and Eq.~\eqref{Schwarzschild-like} are also obtained for the first time. Further more, considering the infinite-distance deflection of light in Schwarzschild-like spacetime, we correct a mistake in Ref.~\cite{Ali181}. By Eq .~\eqref{Schwarzschild-like}, we have 
{\begin{eqnarray}
\label{Schwarzschild-like-infinite}
 \hat{\alpha}_{S_\infty}(v=1)&=&\frac{\pi l}{2}+\frac{4M}{b}+\frac{2Ml}{b}\nn\\
 &&-\frac{\pi l^2}{8}-\frac{M l^2}{2b}+\mathcal{O}(M^2,l^3)~.
\end{eqnarray}}
The first two terms of the expression are consistent with the Ref.~\cite{Casana2018} and Ref.~\cite{Ali181}. However, the third term is different
from the positive and negative signs of the Ref.~\cite{Ali181}.

After obtaining the finite-distance and infinite-distance deflection angles, let us consider their difference described by the finite-distance corrections~\cite{OIA2017,OIA2019}
{\begin{eqnarray}
\label{corrections}
 \delta\hat{\alpha}&=&\hat{\alpha}_{\infty}-\hat{\alpha}~.
\end{eqnarray}}
Bring Eq.~\eqref{Kerrlike} and Eq.~\eqref{massive-infinite} into the above equation, we have
{\begin{eqnarray}
\label{corrections2}
 \delta\hat{\alpha}&=&\left(\lambda-1\right)\left[\arcsin(b u_R)+\arcsin(b u_S)\right]\nn\\
 &&-\bigg[-2\left(1+v^2\right)\lambda\nn\\
 &&+\frac{\left(1+v^2\right)\lambda-b^2 u_R^2\left(1+v^2\lambda\right)}{\sqrt{1-b^2u_R^2}}\nn\\
 &&-\frac{\left(1+v^2\right)\lambda-b^2 u_S^2\left(1+v^2\lambda\right)}{\sqrt{1-b^2u_S^2}}\bigg]\frac{M}{bv^2}\nn\\
 &&\pm\frac{2a M  \lambda\left(2\lambda+\frac{b^2 u_R^2-\lambda}{\sqrt{1-b^2u_R^2}}+\frac{b^2 u_S^2-\lambda}{\sqrt{1-b^2u_S^2}}\right)}{b^2 v}\nn\\
 &&+\mathcal{O}(M^2,a^2)~.
\end{eqnarray}}

\begin{figure}[t]
\begin{center}
\includegraphics[width=0.4\textwidth]{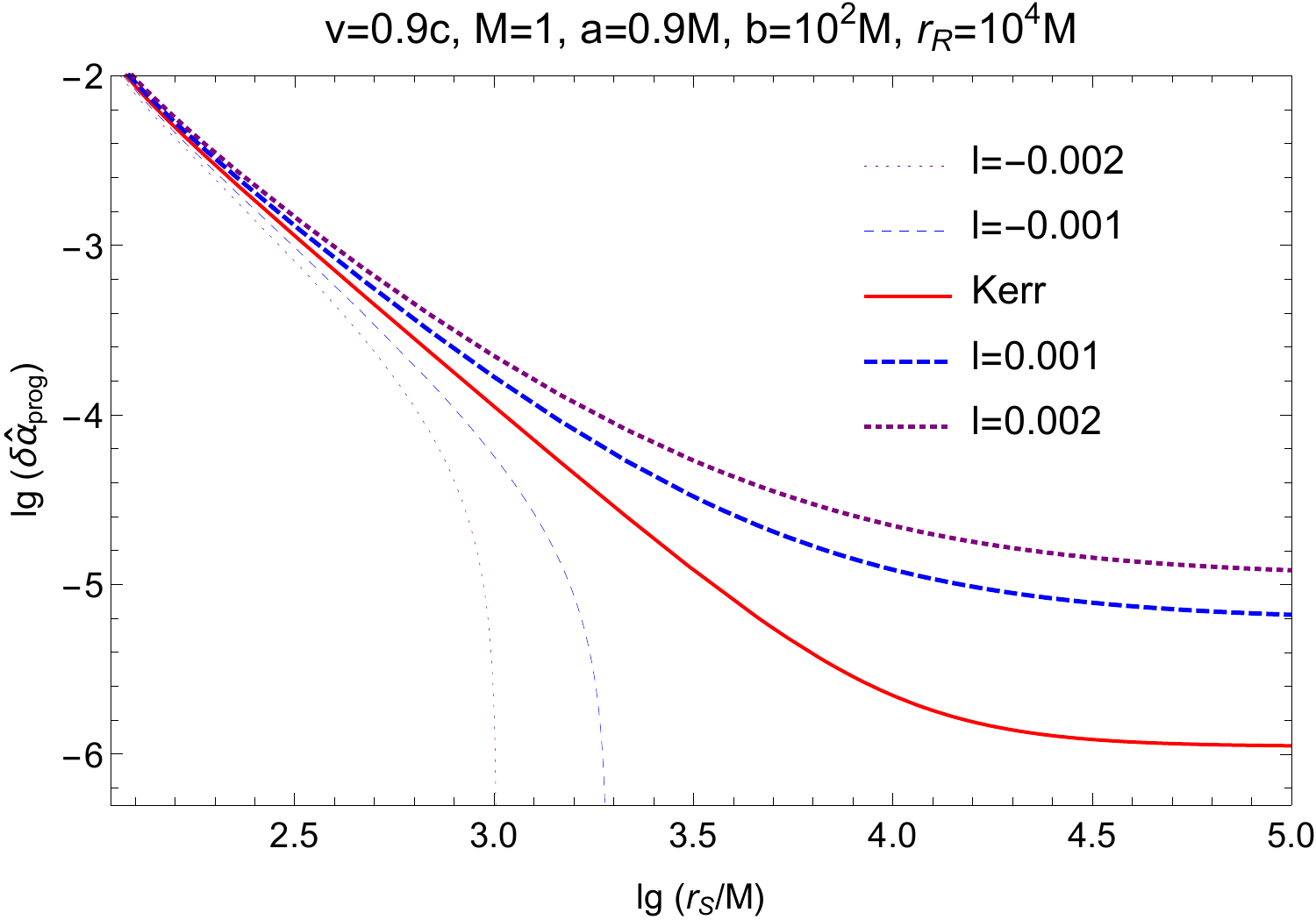}\hspace{0.04\textwidth}\\
(a)\hspace{0.3\textwidth}\\
\includegraphics[width=0.4\textwidth]{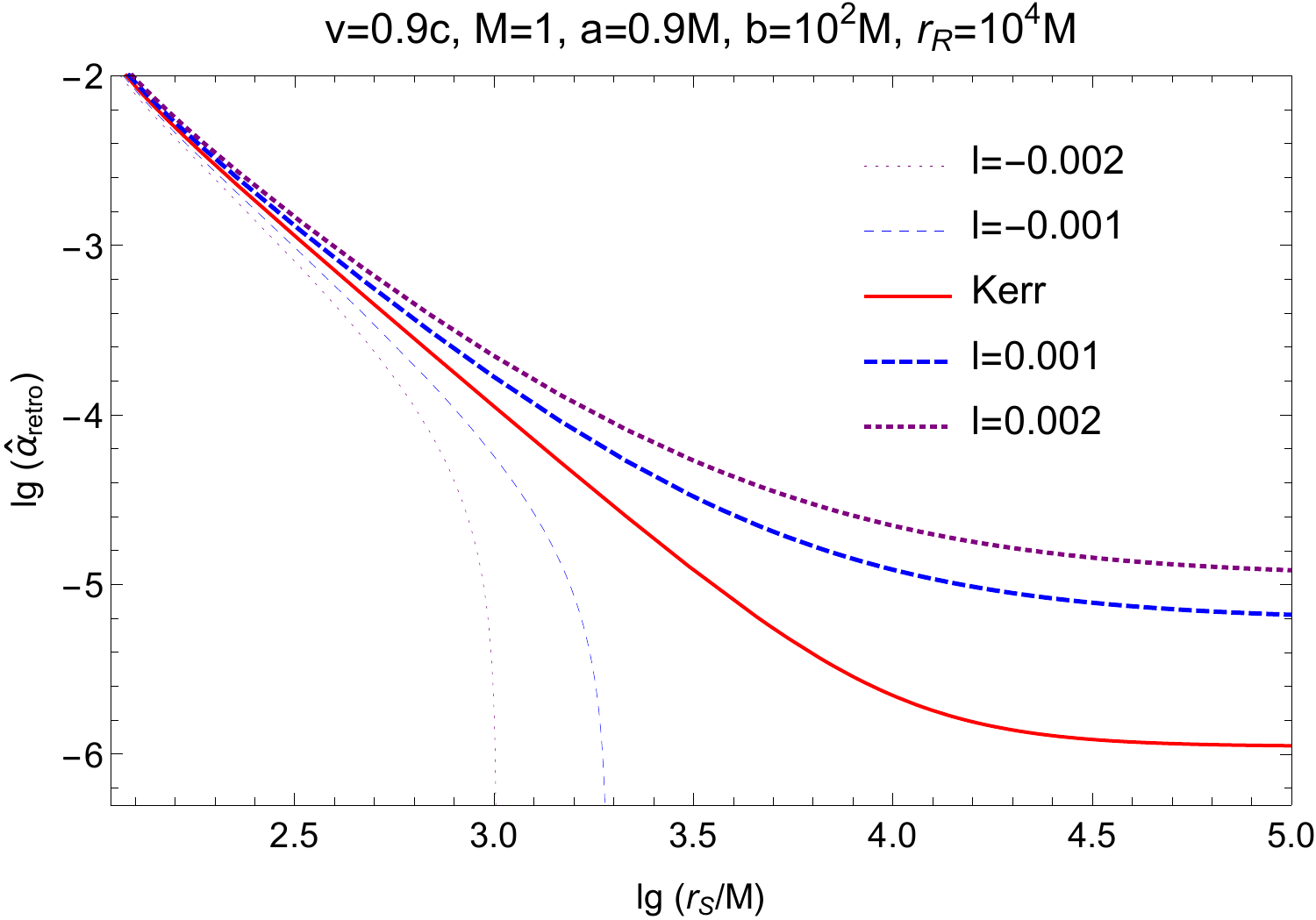}\hspace{0.04\textwidth}\\
(b)\hspace{0.3\textwidth}\\
\caption{The finite-distance corrections in Eq.~\eqref{corrections2}.  The vertical axis denotes $\lg\hat{\alpha}$, and the horizontal axis denotes $\lg~(r_S/M)$. (a) denotes the prograde particle deflection and (b) denotes the retrograde particle deflection.} \label{figcorreangle}
\end{center}
\end{figure}

In Fig.~\ref{figcorreangle}, we use the same parameters as in Fig.~\ref{figangle} and plot the finite-distance corrections in Eq.~\eqref{corrections2}, where the vertical axis denotes $\lg\hat{\alpha}$, and the horizontal axis denotes $\lg~(r_S/M)$. We can see that the effect of the Lorentz breaking constant $l$ on the finite-distance correction is similar to the deflection angle when the receive distance $r_R$ and source distance $r_R$ are not too large.

In addition, we can use $bu_R$ and $bu_S$ to expand~\eqref{corrections2}, and the result reads
{\begin{eqnarray}
\label{corrections3}
 \delta\hat{\alpha}&=&b\left(u_R+u_S\right)\left(\lambda-1\right)\nn\\
 &&-\frac{b M\left(u_R^2+u_S^2\right)\left(2-\lambda+v^2\lambda\right)}{2v^2}\nn\\
 &&\pm\frac{a M\left(u_R^2+u_S^2\right)\left(2-\lambda\right)\lambda} {v}\nn\\
&&+\mathcal{O}(M^2,a^2,b^3u_R^3,b^3u_S^3)~.
\end{eqnarray}}
It is interesting to point out that the term including $a M$ is independent of the impact parameter.

\section{conclusion} \label{CONCLU}

In the weak-field limits, we have studied the deflection angle of massive particles for an observer and source at a finite distance from  the Kerr-like black hole in the bumblebee gravity theory. For this purpose, we used a geometric and topological approach first proposed by Gibbons and Werner~\cite{GW2008,Werner2012} and then extended by Ono~\textit{et al} to study the finite-distance deflection of light~\cite{OIA2017,OIA2018,OIA2019}. This method involves the application of the Gauss-Bonnet theorem. To apply this method, we used the Jacobi-Maupertuis Randers-Finsler metric. Furthermore, since the spacetime is asymptotically non-flat the calculation of the deflection angle is modified in Eq.~\eqref{gbdef}. In addition, we directly computed the deflection angle using its definition proposed in Refs.~\cite{OIA2017,OIA2018,OIA2019}. The results obtained by the two methods are the same, and are shown in Eq.~\eqref{gbdef}. Our results can also deduce some new results; for example, the finite-distance deflection angle of light in Eq.~\eqref{light-finite}, the infinite-distance deflection angle of a massive particle in Eq.~\eqref{massive-infinite} and the finite-distance deflection angle of a massive particle by a Schwarzschild-like black hole in the bumblebee gravity theory in Eq.~\eqref{Schwarzschild-like}. In particular, when considering the limit of deflection of light in Schwarzschild-like spacetime, in Eq.~\eqref{Schwarzschild-like-infinite} we correct a mistake in Ref.~\cite{Ali181} . Furthermore, we also obtained the finite-distance correlation of the gravitational deflection angle of massive particles.

From our result, it show that the deviation from general relativity: when Lorentz breaking constant $l>0$, the deflection angle is larger than it by Kerr black hole ($l=0$); when $l<0$, the deflection angle is smaller than it by Kerr black hole. On the whole, the deflection increases as $l$ increases. Interestingly, the same relationship applies to finite-distance correlations when receive distance $r_R$ and source distance $r_R$ are not too large, as shown in Fig.~\ref{figcorreangle}.
 
  \acknowledgments
This work was supported by Comisi{\'o}n Nacional de Ciencias y Tecnolog{\'i}a of Chile through FONDECYT Grant $N^\mathrm{o}$ 3170035 (A. {\"O}.).

\appendix
\section{The gravitational deflection angle of massive particles using Werner's method} \label{AppendixA}
In this appendix, we shall compute the infinite-distance gravitational deflection angle of massive particles by Kerr-like black hole in bumblebee gravity model using Werner's Finsler geometry method~\cite{Werner2012}. The Hessian of Finsler metric $F(x,y)$ reads~\cite{Chern2002}
{\begin{eqnarray}
\label{Hessian}
 g_{ij}(x,y)&=&\frac{1}{2}\frac{\partial^2F^2(x,y)}{\partial y^i \partial y^j}~,
\end{eqnarray}}
where $(x,y)\in TM$ with $TM$ being the tangent bundle of smooth manifold $M$.
In Ref.~\cite{Werner2012}, Werner applied Naz{\i}m's method to construct an  osculating Riemannian manifold $(M, \tilde{g})$ of Finsler manifold $(M, F)$. Following Werner, we can choose a smooth nonzero vector field $Y$ tangent to the geodesic $\gamma_F$, i.e. $Y(\gamma_F)=y$, and thus the osculating Riemannian metric can be obtained by Hessian
{\begin{eqnarray}
\label{Hessian1}
 \tilde{g}_{ij}(x)&=&g_{ij}\left(x,Y(x)\right)~.
\end{eqnarray}}
In this construction, the geodesic in $(M, F)$ is also a geodesic in $(M, \tilde{g})$~\cite{Werner2012}.
 On the equatorial plane, our Randers-Finsler metric in Eq.~\eqref{KerrJMRE} leads to
\begin{widetext}
{\begin{eqnarray}
\label{Randers-FinslerY}
 F\left(r,\varphi,Y^r,Y^\varphi\right)&=&\sqrt{\mathcal{E}^2\left(\frac{1}{1-\frac{2M}{r}}-1+v^2\right)\bigg[\frac{\lambda^2r^2}{a^2\lambda^2+r\left(r-2M\right)}\left(Y^r\right)^2+\frac{r\left(a^2\lambda^2+r^2-2Mr\right)}{r-2M}\left(Y^\varphi\right)^2\bigg]}\nn\\
 &&-\frac{2\lambda\mathcal{E} M a r}{r(r-2M)}Y^\varphi~.
\end{eqnarray}}

In order to obtain the leading-order deflection angle, we only need the zeroth-order particle ray $r=b/{\sin(\frac{\varphi}{\lambda})}$. Near the particle ray, one can choose the vector field as follows
{\begin{eqnarray}
\label{vector field}
Y^r=\frac{dr}{d\sigma}=-\frac{\cos(\frac{\varphi}{\lambda})}{\mathcal{E} \lambda v}~,~~~Y^\varphi=\frac{d\varphi}{d\sigma}=\frac{\sin^2(\frac{\varphi}{\lambda})}{\mathcal{E} b v}~.
\end{eqnarray}}
According Eqs.~\eqref{Hessian}-~\eqref{vector field}, we can get the osculating Riemannian metric as following
{\begin{eqnarray}
\tilde{g}_{rr}&=&\mathcal{E}^2 v^2 \lambda^2+\frac{2\mathcal{E} M\left(1+v^2\right)\lambda^2}{r}-\frac{2a M \mathcal{E}^2 r v \lambda^3 \sin^6\left(\frac{\varphi}{\lambda}\right)}{b^3\left[\cos^2\left(\frac{\varphi}{\lambda}\right)+\frac{r^2}{b^2}\sin^4\left(\frac{\varphi}{\lambda}\right)\right]^{3/2}}+\mathcal{O}\left(M^2,a^2\right)~,\\
\tilde{g}_{r\varphi}&=&-\frac{2a M \mathcal{E}^2 v \lambda^2 \cos^3\left(\frac{\varphi}{\lambda}\right)}{r\left[\cos^2\left(\frac{\varphi}{\lambda}\right)+\frac{r^2}{b^2}\sin^4\left(\frac{\varphi}{\lambda}\right)\right]^{3/2}}+\mathcal{O}\left(M^2,a^2\right)~,\\
\tilde{g}_{\varphi\varphi}&=&\mathcal{E}^2 v^2 r^2+2\mathcal{E}^2 M r-\frac{2a M \mathcal{E}^2 r v \lambda \sin^2\left(\frac{\varphi}{\lambda}\right)\left[3\cos^2(\frac{\varphi}{\lambda})+2\frac{r^2}{b^2}\sin^4(\frac{\varphi}{\lambda})\right]}{b\left[\cos^2\left(\frac{\varphi}{\lambda}\right)+\frac{r^2}{b^2}\sin^4\left(\frac{\varphi}{\lambda}\right)\right]^{3/2}}+\mathcal{O}\left(M^2,a^2\right)~.
\end{eqnarray}}
Then, the corresponding Gaussian curvature can be compute by Eq.~\eqref{Gauss-K} and the result up to leading order is
{\begin{eqnarray}
\tilde{K}&=&-\frac{\left(1+v^2\right)M}{\mathcal{E}^2 r^3v^4\lambda^2}+\frac{3a M}{\mathcal{E}^2 v^3 \lambda b^2 r^2} f\left(r,\varphi\right)+\mathcal{O}\left(M^2,a^2\right)~,
\end{eqnarray}}
where
{\begin{eqnarray}
f\left(r,\varphi\right)&=&\frac{\sin^3(\frac{\varphi}{\lambda})}{\left[\cos^2\left(\frac{\varphi}{\lambda}\right)+\frac{r^2}{b^2}\sin^4\left(\frac{\varphi}{\lambda}\right)\right]^{7/2}}\Bigg\{2\cos^6\left(\frac{\varphi}{\lambda}\right)\left[-2+5\frac{r}{b}\sin\left(\frac{\varphi}{\lambda}\right)\right]\nn\\
&&+2\frac{r}{b}\cos^2(\frac{\varphi}{\lambda})\sin^5(\frac{\varphi}{\lambda})\left[2-\frac{r^2}{b^2}+\frac{r^2}{b^2}\cos\left(\frac{2\varphi}{\lambda}\right)+4\frac{r}{b}\sin\left(\frac{\varphi}{\lambda}\right)\right]\nn\\
&&+\cos^4\left(\frac{\varphi}{\lambda}\right)\sin^2\left(\frac{\varphi}{\lambda}\right)\left[-2+9\frac{r}{b}\sin\left(\frac{\varphi}{\lambda}\right)-10\frac{r^3}{b^3}\sin^3\left(\frac{\varphi}{\lambda}\right)\right]\nn\\
&&+\frac{r^2}{b^2}\left[-\frac{r}{b}\sin^9\left(\frac{\varphi}{\lambda}\right)+2\frac{r^3}{b^3}\sin^{11}\left(\frac{\varphi}{\lambda}\right)+\sin^4\left(\frac{2\varphi}{\lambda}\right)\right]\Bigg\}~.
\end{eqnarray}}
In infinite-distance case, we consider the area enclosed by particle lines and curve $C_\infty$ instead of quadrilateral $\prescript{\infty}{R}\Box_{S}^{\infty}$. Notice that now the particle ray is a geodesic in $(M,\tilde{g})$ and thus $k_g=0$. Then by using Gauss-Bonnet theorem, the infinite-distance deflection angle can be calculated by
{\begin{eqnarray}
\hat{\alpha}_{\infty}&=&\left(\lambda-1\right)\pi-\lambda\int\int_\infty \tilde{K} dS ~.
\end{eqnarray}}
Then, the infinite-distance deflection angle of massive particles can be calculated by
{\begin{eqnarray}
\hat{\alpha}_{\infty}&=&\left(\lambda-1\right)\pi-\lambda\int_0^{\lambda\pi}\int_{b/{\sin(\frac{\varphi}{\lambda})}}^\infty \tilde{K}\sqrt{det\tilde{g}}dr d\varphi~\nn\\
&=&\left(\lambda-1\right)\pi+\int_0^{\lambda\pi}\int_{b/{\sin(\frac{\varphi}{\lambda})}}^\infty \left[\frac{\left(1+v^2\right)M}{ r^2v^2}-\frac{3a M \lambda}{ b^2 v r } f\left(r,\varphi\right)+\mathcal{O}\left(M^2,a^2\right)\right]dr d\varphi~\nn\\
&=&\left(\lambda-1\right)\pi+\frac{2M \left(1+v^2\right)\lambda}{bv^2}-\frac{4a M \lambda^2}{b^2v}+\mathcal{O}(M^2,a^2)~.
\end{eqnarray}}
\end{widetext}

\end{document}